\documentclass[11pt]{article}
\usepackage[bottom=20mm]{geometry}
\usepackage{color}

\usepackage{latexsym}  
\usepackage{amssymb}
\usepackage{placeins} 
\usepackage{graphicx,xcolor}
\usepackage{amsmath}
\usepackage{mathrsfs}
\usepackage{caption}
\newcommand{\be}{\begin{equation}}
\newcommand{\ee}{\end{equation}}
\newcommand{\bea}{\begin{eqnarray}}
\newcommand{\eea}{\end{eqnarray}}
\newcommand{\bref}[1]{(\ref{#1})}

\newcommand{\bi}{\bibitem}

\begin{document}
\begin{titlepage}
\begin{flushright}
\today
\end{flushright}

\begin{center}
{\Large\bf  
Axion in the minimal SO(10) GUT}
\end{center}

\begin{center}

\vspace{0.1cm}

{\large Takeshi Fukuyama%
\footnote{E-mail: fukuyama@rcnp.osaka-u.ac.jp}}

\vspace{0.2cm}

{\small \it Research Center for Nuclear Physics (RCNP),
Osaka University, \\Ibaraki, Osaka, 567-0047, Japan}


\end{center}

\begin{abstract}
The QCD axion is investigated within the minimal supersymmetric SO(10) grand unified theory, where the Yukawa sector involves Higgs multiplets ${\bf 10}$ and $\overline{{\bf 126}}$.
The relative phase between the VEVs of $({\bf 10,1,3})\subset\overline{{\bf 126}}$ and $({\bf \overline{10},1,3})\subset{\bf 126}$ under ${\rm SU}(4)_C\times{\rm SU}(2)_L\times{\rm SU}(2)_R$ is identified with the axion.
The Peccei–Quinn and $B-L$ symmetry breaking scales coincide through $|\langle\Delta_R\rangle|=|\langle\overline{\Delta}_R\rangle|$.
The scalar partner of the lightest right-handed neutrino plays the role of the inflaton, realizing hybrid inflation consistent with the observed CMB density fluctuations.
After inflation, both fields acquire VEVs, and the domain-wall problem is resolved through the Lazarides–Shafi mechanism, which naturally restricts the model to three generations.

\end{abstract}
Key words: axion; SO(10) GUT; sneutrino inflaton
\end{titlepage}

The axion is regarded as a leading dark matter candidate, owing to its well-defined role in preventing the violation of strong CP symmetry \cite{PQ1, PQ2, WW1, WW2, KSVZ1, KSVZ2, ZDFS1, ZDFS2}.
However, there are many discussions concerning axions and axion-like particles. In this letter, we restrict our focus to the QCD axion.
Even within the framework of the QCD axion, two characteristic energy scales typically appear: the conventional QCD axion and the heterotic string–inspired axion.
The former corresponds to a bottom-up approach, whereas the latter emerges from a top-down framework. Although both are commonly referred to as “axions,” they are associated with fundamentally different energy scales.
 In this letter, we consider the top-down scenario within the framework of the conventional QCD axion. By examining the same axion from both perspectives, we aim to obtain more stringent and fruitful insights. The main point of this letter is as follows. The PQ and $B-L$ symmetries are broken after the end of inflation. It is noteworthy that the axion field can be naturally incorporated within the minimal SO(10) model, in which the PQ symmetry and the right-handed neutrino sector are intimately connected. This unified framework provides a consistent explanation linking the origin of neutrino masses, the strong CP problem, and the post-inflationary dynamics of symmetry breaking.

Let us begin with the review of the conventional bottom-up approach:
Axion field ($a(x)$) potential in the classic dilute gas approximation becomes
\be
V(a(x))=m_u\Lambda_{QCD}^3\left[1-\cos\left(\frac{a(x)}{f_a}\right)\right]\equiv \Lambda^4\left[1-\cos\left(\frac{a}{f_a}\right)\right] \;.
\label{potential}
\ee
Using the Gellmann-Oakes-Renner relation \cite{Gellmann}, 
\be
\Lambda_{QCD}^3=-<\overline{q}q>=\frac{F_\pi^2m_\pi^2}{m_u+m_d} \;.
\label{LQCD}
\ee
Thus, it means that a massive pion appears as a pseudo Nambu-Goldstone particle from the chiral symmetry breaking.
Here $F_\pi=93$ MeV, and $\frac{m_u}{m_d}\approx 0.47$, and we obtain axion mass
\be
m_a=5.7\times 10^{-6}\left(\frac{10^{12}\mbox{GeV}}{f_a}\right) \text{eV} \;.
\label{ma}
\ee
Then, the axion self coupling constant $\lambda$ in $\frac{\lambda}{4!}a(x)^4$ becomes
\be
\lambda=-0.47\frac{F_\pi^2m_\pi^2}{f_a^4}<0 \;.
\label{selfc}
\ee
Here $f_a$ is not determined intrinsically but restricted from astrophysical and cosmological bounds.
Constraints from the energy loss of super novae \cite{Kim, Raffelt, Janka}
\be
f_a>10^9~\mbox{GeV} \;.
\ee
Constraints from relic density of DM $\Omega_c$ \cite{Preskill, Sikivie, Dine}
\be
\Omega_ah^2\approx 0.1\theta_i^2\left(\frac{f_a}{10^{12}\mbox{GeV}}\right)^{7/6}\leq \Omega_ch^2=0.120\pm0.001 \;,
\ee
where $\theta_i$ is the initial misalignment angle. 

On the other hand, heterotic superstring theory gives on the QCD axion the different form, 
\be
V(\phi)=\Lambda_{string}^4\left[1-\cos\left(\frac{a(x)}{f_a}\right)\right]
\ee
with
\be
\Lambda_{string}^4=M_{SUSY}^2M_{Pl}^{*2}e^{-S_{instanton}} 
\label{Lstring}
\ee
and $f_a$ is given by \cite{Svrcek}
\be
f_a=\frac{\alpha_{GUT}M_{Pl}^*}{\sqrt{2}2\pi}\approx 1.1\times 10^{16}~~\mbox{GeV} \;.
\label{fa}
\ee
Here we have set the unified coupling constant 
\be
\alpha_{GUT}=\frac{1}{25} \;.
\ee
As for the axion mass $m_a$, it depends crucially on the instanton action $S_{instanton}$ \cite{Svrcek, Visinelli, Fuku7}.
Thus, Eq.~\bref{LQCD} and Eq.~\bref{Lstring} may be regarded as bottom–up and top–down approaches, respectively.
However, the underlying dynamics and energy scales in these two approaches are quite different.
In this letter, we focus on the top–down scenario of the conventional axion \bref{LQCD} within the framework of the SO(10) GUT. 
There exist many works in this scheme \cite{Lazarides, Reiss, Kikuchi, Ernst, Boucenna, Hamada, Babu2}.  In this letter we revisit axion in the frame work of the minimal SO(10) GUT model with two Higgs of ${\bf 10}$ and ${\bf \overline{126}}$ dimensional Higgs scalars \cite{Kikuchi} in the matter coupling. The details of this model is given in \cite{Fuku1,Fuku11, Babu, Fuku2,Fuku3}. Here the symmetry breaking pattern is given by
\be
SO(10)\rightarrow SU(4)_C\times SU(2)_L\times SU(2)_R\rightarrow SU(3)_C\times SU(2)_L\times U(1)_Y \;.
\ee

In order to avoid a heavy axion mass, which does not solve the strong CP 
problem, we also impose a discrete symmetry ${\mathbb Z}_3$. 
The corresponding charges with regards to this ${\mathbb Z}_3$ symmetry are 
listed in Table 1. Then, the ${\rm SO}(10) \times {\mathbb Z}_3$ invariant 
superpotential is given by 
\be
W = \Psi_i (Y_{10}^{ij}\; H 
+ Y_{126}^{ij} \;{\bf \bar{\Delta}} ) \Psi_j 
+ m_1 {\bf \overline{\Delta}} {\bf \Delta}
+ m_2 \Phi^2
+ \lambda_1 {\bf \bar{\Delta}} {\bf \Delta} \Phi
+ \lambda_2 {\bf \Delta} H \Phi 
+ \lambda_3 \Phi^3 \;, 
\label{W1}
\ee
where $\Psi_i$ is a ${\bf 16}$-dimensional matter multiplet, 
$H$ is a ${\bf 10}$-dimensional multiplet which essentially gives 
a large top Yukawa coupling and $\overline{\Delta}~(\Delta)$ is $\overline{{\bf 126}}~({\bf 126})$-dimensional multiplet. $\Phi$ is a ${\bf 210}$-dimensional 
multiplet that is used to break the SO(10) gauge symmetry. 
The details of this potential can be found in 
\cite{Fuku3, Fuku4}. 
The Pati-Salam phase is broken by the vev of $\overline{\Delta}_R=({\bf 10,1,3})$, and the right handed neutrino masses are generated through the following type of Yukawa 
interaction, 
\be
W = Y_{126}^{ij} ~{\bf \bar{\Delta}}_R \,
\nu_{Ri}^T C^{-1} \nu_{Rj} \;.
\ee
This gives the Majorana masses for the right-handed neutrinos, 
\be
M_R^{ij} = Y_{126}^{ij} \left<{\bf \bar{\Delta}}_R \right> \;.  
\ee
Here we list some results necessary for the following arguments:
$Y_\nu$ is the neutrino Dirac Yukawa coupling matrix, and determined as \cite{Fuku2}
\begin{eqnarray}
 Y_{\nu} = 
\left( 
 \begin{array}{ccc}
-0.000135 - 0.00273 i & 0.00113  + 0.0136 i  & 0.0339   + 0.0580 i  \\ 
 0.00759  + 0.0119 i  & -0.0270   - 0.00419  i  & -0.272    - 0.175   i  \\ 
-0.0280   + 0.00397 i & 0.0635   - 0.0119 i  &  0.491  - 0.526 i 
 \end{array}   \right) 
\label{Ynu}
\end{eqnarray}     
and heavy right-handed masses are
\be
M_{R1}\approx 1.2\times 10^{11}~\mbox{GeV},~~M_{R2}\approx 1.8\times 10^{12}~\mbox{GeV},~~M_{R3}\approx 8.3\times 10^{12}~\mbox{GeV} \;.
\label{MR}
\ee
However, even within the minimal SO(10) framework, certain ambiguities remain in the numerical values. The possible ranges of these parameters are given in Ref.~\cite{Fuku6}, and they will be discussed in detail later in this Letter.

The essential point in this framework to generate the PQ axion is as follows: 
the ${\bf \overline{126}}$ and the ${\bf 126}$ are independent fields required 
in order to preserve SUSY, but they always appear in pairs, and the SUSY 
vacuum condition (D-flat condition) can never determine the relative phase 
degree of freedom: 
\be
\left|\left<{\bf \bar{\Delta}}_R \right>\right|^2 - 
\left|\left< {\bf \Delta}_R  \right> \right|^2 =0 \;. 
\ee
Here $\Delta_R=(10,1,3)$ under Pati-Salam phase $SU(4)\times SU(2)_L\times SU(2)_R$.
$<\Delta_R>\neq 0$ induces
\be
SU(2)_R\times U(1)_{B-L}\rightarrow U(1)_Y \;.
\label{B-L}
\ee
This means, the relative phase remains as a physical degree of freedom, 
the so called pseudo-NG boson. Schematically, we can write this fact as 
follows: 
\be
\left< {\bf \bar{\Delta}}_R \right> \sim 
\left< {\bf \Delta}_R \right> \times \exp(i\, \Theta) \;, 
\ee
where the argument field or the pseudo-NG boson $\Theta=\frac{a}{f_a}$ can be 
regarded as the axion \cite{Kikuchi}. It gives a connection between the $U(1)_{B-L}$ 
symmetry breaking scale $(\Lambda_{B-L})$ and the $U(1)_{\rm PQ}$ 
symmetry breaking scale ($\Lambda_{\rm PQ}$)\footnote{Note that since these two fields ${\bf \bar{\Delta}}_R$ and ${\bf \Delta}_R$ 
are completely independent, one of which is used to break the $B-L$ symmetry 
and the other can be used to break the PQ symmetry as well. Remarkably, 
the former symmetry is gauged in SO(10) although the latter one is ungauged, 
hence one of the NG bosons residing in the above fields is absorbed into 
the $B-L$ gauge boson, but the other remains as a physical degree of freedom, 
the axion. }.
That is one of our main conclusions in this article. 
In general, we can assign a global $U(1)_{PQ}$ charge to these fields. 
For instance, ${\rm PQ}[{\bf \bar{\Delta}}_R] = +2$, ${\rm PQ}[\nu_{R}] = -1$. 
Then after giving rise to the VEV of ${\bf \bar{\Delta}}_R$, 
the global $U(1)_{\rm PQ}$ symmetry would be spontaneously broken and 
there appears a pseudo-NG boson that is later understood as the axion. 
The scalar potential of the ${\bf \bar{\Delta}}_R$ field 
includes the mixing term with the electroweak Higgs doublets, 
\be
V = \lambda~ {\bf \bar{\Delta}}_R {\bf \Delta}_R H_{126} H_{10}\;. 
\ee
Here $H_{10} \equiv ({\bf 1,2,2})$ and $H_{126} \equiv ({\bf 15,2,2})$ 
are the $SU(2)_L$ bi-doublet Higgs fields arising from the ${\bf 10}$ and 
${\bf \overline{126}}$ multiplets of SO(10), respectively, and 
${\bf \Delta}_R$ is required for the anomaly cancellation. 
A linear combination of 
$H_{10}$ and $H_{126}$ Higgs fields becomes the MSSM Higgs doublets 
$H_u$ and $H_d$ that cause the correct electroweak symmetry breaking, 
$SU(2)_L \times U(1)_Y \to U(1)_{\rm em}$. 
This potential would cause a connection between intermediate scale physics 
and the electroweak scale. Since the fields $H_u$ and $H_d$, 
or equivalently $H_{10}$ and $H_{126}$ have couplings to the quarks 
and leptons, a rotation of the Higgs fields $H_{10} \to H_{10} 
\exp(+ 2 i \theta)$ and $H_{126} \to H_{126} \exp(+2 i \theta)$ gives 
a chiral rotation of the quarks and leptons 
$\{q_L, u_R^c, d_R^c, \ell_L, \nu_R^c, e_R^c \} \to \exp(- i\theta) 
\{q_L, u_R^c, d_R^c, \ell_L, \nu_R^c, e_R^c \}$. 
Such a non-trivial transformation indicates an anomalous symmetry 
and it induces an anomalous coupling of the pseudo-NG boson $a(x)$ 
to the gluon field. 
\be
{\cal L}= \frac{a(x)}{f_a} \,
\frac{g_s^2}{32 \pi^2}\, G_{\mu \nu}^A \tilde{G}_A^{\mu \nu}\;, 
\ee
where $g_s$ is the $SU(3)_c$ gauge coupling constant, 
$G_{\mu \nu}^A$ is the gluon field strength and $\tilde{G}^{\mu\nu}_A 
\equiv \frac{1}{2} \epsilon^{\mu \nu \rho \sigma} G_{\rho \sigma}^A$. 
This kind of interaction is used to solve the strong CP problem. 
Then the interaction of the axion with the quarks and leptons is given by 
\be
{\cal L} \ =\ \frac{a(x)}{f_a} {\partial}_{\mu} J^{\mu}, 
\ee
where $J^{\mu}$ is a conserved current 
associated with the global $U(1)_{\rm PQ}$ symmetry 
\be
J^{\mu}
= f_a {\partial}^{\mu} a(x) 
+ 2 \sin^2\beta \;\bar{u}_i \gamma^\mu \gamma_5 u_i 
+ 2 \cos^2\beta \;\bar{d}_i \gamma^\mu \gamma_5 d_i 
+ 2 \cos^2\beta \;\bar{e}_i \gamma^\mu \gamma_5 e_i   
\ee
with $\tan \beta \equiv \left<H_u \right>/\left<H_d \right>$. 
As is usual for the pseudo-NG bosons, the mass of the axion 
is inversely proportional to the decay constant $f_a$ as 
\be
m_a = 0.62  \times 10^{-6} \;{\rm eV} \; \times
\frac{10^{13} \;{\rm GeV}}{f_a} \;.
\ee


After the SO(10) symmetry breaking, we have the following 
superpotential for the matter multiplets: 
\bea
W &=& u_{Ri}^c 
\left(Y_{10}^{ij} H_{10}^u + Y_{126}^{ij} H_{126}^u \right) q_{Lj}
\ +\ d_{Ri}^c 
\left(Y_{10}^{ij} H_{10}^d + Y_{126}^{ij} H_{126}^d \right) q_{Lj}
\nonumber\\
&+& N_{i}^c 
\left(Y_{10}^{ij} H_{10}^u -3\,Y_{126}^{ij} H_{126}^u \right) \ell_{Lj}
\ +\ e_{Ri}^c 
\left(Y_{10}^{ij} H_{10}^d -3\,Y_{126}^{ij} H_{126}^d \right) \ell_{Lj}
\nonumber\\
&+& Y_{126}^{ij} \,N_i^c N_j^c \,{\bf \bar{\Delta}}_R \;.
\eea
Each field can have the PQ charges as listed in Table 1. 
\begin{table}
\begin{center}
\begin{tabular}{|c|c|c|}
\hline \hline
fields & PQ charges & ${\mathbb Z}_3$ charges \\
\hline
$\Psi $ & $-1$ & $\omega^2$ \\
$H$ & $+2$ & $\omega^2$ \\
${\bf \bar{\Delta}}$ & $+2$ & $\omega^2$ \\
${\bf \Delta}$ & $-2$ & $\omega$ \\
$\Phi$ & $0$ & $1$ \\
\hline \hline
\end{tabular}
\caption{PQ and ${\mathbb Z}_3$ charges of the fields ($\omega^3 =1$). }
\end{center}
\end{table}
In addition to this, we have the soft SUSY breaking terms defined as 
follows: 
\be
V_{\rm SOFT} =
m_{\tilde{N}_i}^2 \; |\tilde{N}_i|^2
+ m_{\bar{\Delta}}^2 \; \left|{\bf \bar{\Delta}}_R \right|^2 
+ m_{\Delta}^2 \; \left|{\bf \Delta}_R \right|^2 
+ \left(A_{N}^{ij} \; {\bf \bar{\Delta}}_R \; \tilde{N}_{i} 
\tilde{N}_{j} \ +\ {\rm H.c.} \right) \;,
\ee
where $A_N^{ij}$ is the tri-linear coupling constant which is 
assumed to be proportional to the Yukawa coupling constant $Y_{126}^{ij}$, 
$A_N^{ij} = m_{3/2} Y_{126}^{ij}$. From the superpotential given above, 
we can calculate the scalar potential in the usual way: 
\be
V=
\left|\frac{\partial{W}}{\partial{{\bf \bar{\Delta}}_R}} \right|^2
\ +\ \left|\frac{\partial{W}}{\partial{\tilde{N}}} \right|^2 
\ +\ V_{\rm SOFT} \;,
\ee
that is, 
\be
V=m_{\tilde{N}_i}^2 \, |\tilde{N}_i|^2
+ \left(M_{\rm GUT}^2 + m_{\bar{\Delta}}^2 \right)\, 
\left| {\bf \bar{\Delta}}_R \right|^2
+ \left\{
\left(Y_{126}^{ij} M_{\rm GUT} + A_N^{ij} \right)\, 
{\bf \bar{\Delta}}_R \; \tilde{N}_i^* \tilde{N}_j + {\rm H.c.} \right\} 
 + \cdots \;.
\label{VPQ}
\ee
We regard the scalar partner of the lightest right-handed neutrino 
(sneutrino) $\tilde{N}_1$ as the inflaton, 
that is, we consider the sneutrino inflation scenario \cite{Kikuchi, sneutrino, Ellis, Kallosh, Antusch} .
The reasons for selecting the lightest $\tilde{N}_1$ as the sneutrino inflaton are as follows:
\begin{enumerate}
\item The radiative corrections cause only a small violation of potential flatness.
\item The reheating temperature remains low.
\item The supergravity (SUGRA) corrections are minimal.
\end{enumerate}

\begin{figure}[tbp]
  \begin{center}
   \includegraphics[width=0.5\linewidth,keepaspectratio]{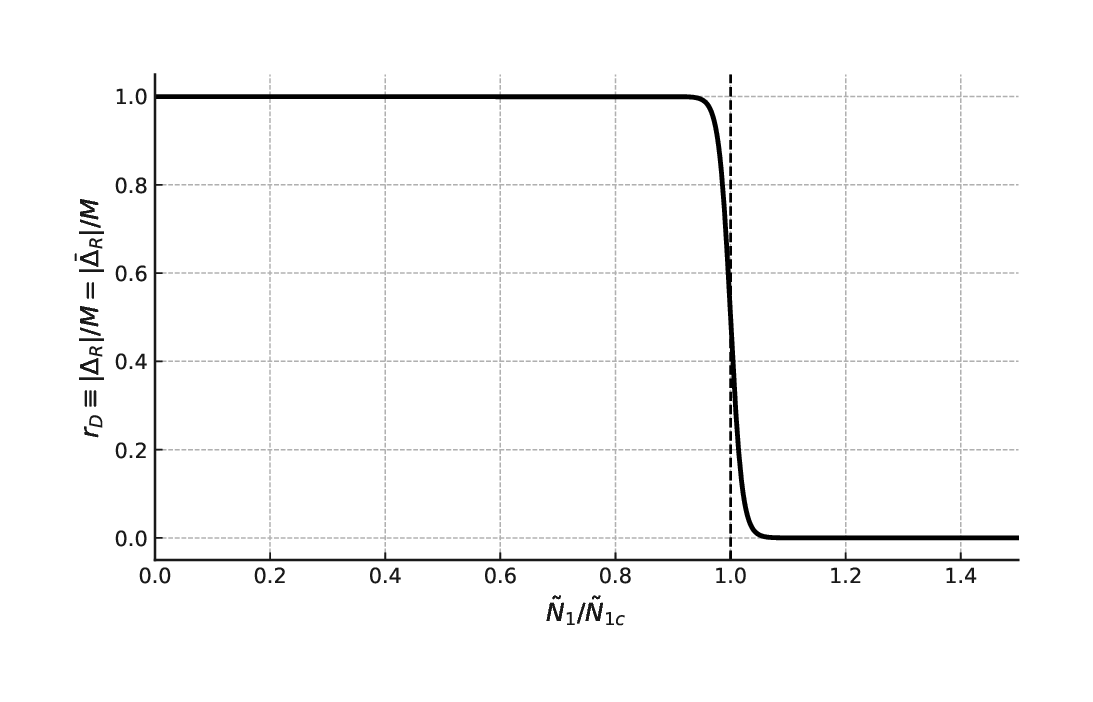}
 \caption{Hybrid-inflation waterfall schematic: Order parameter $r_D\equiv |\Delta_R|/M=|\overline{\Delta}_R|/M$ as a function of the inflaton ratio $\tilde{N}_1/\tilde{N}_{1c}$ in F-term hybrid inflation \cite{Linde, Dvali}. For $\tilde{N}_1>\tilde{N}_{1c}$ (right of the dashed line) the waterfall fields are stabilized at the origin and $r_D=0$ (inflation phase). When $\tilde{N}_1$ reaches the critical value $\tilde{N}_{1c}$, $m_\Delta^2$ turns tachyonic and the system rolls to the symmetry-breaking phase. Time flows from right to left along the horizontal axis.}
  \end{center}
  \label{fig:waterfall1}
\end{figure}
\begin{figure}[h]
  \begin{center}
   \includegraphics[width=0.9\linewidth,keepaspectratio]{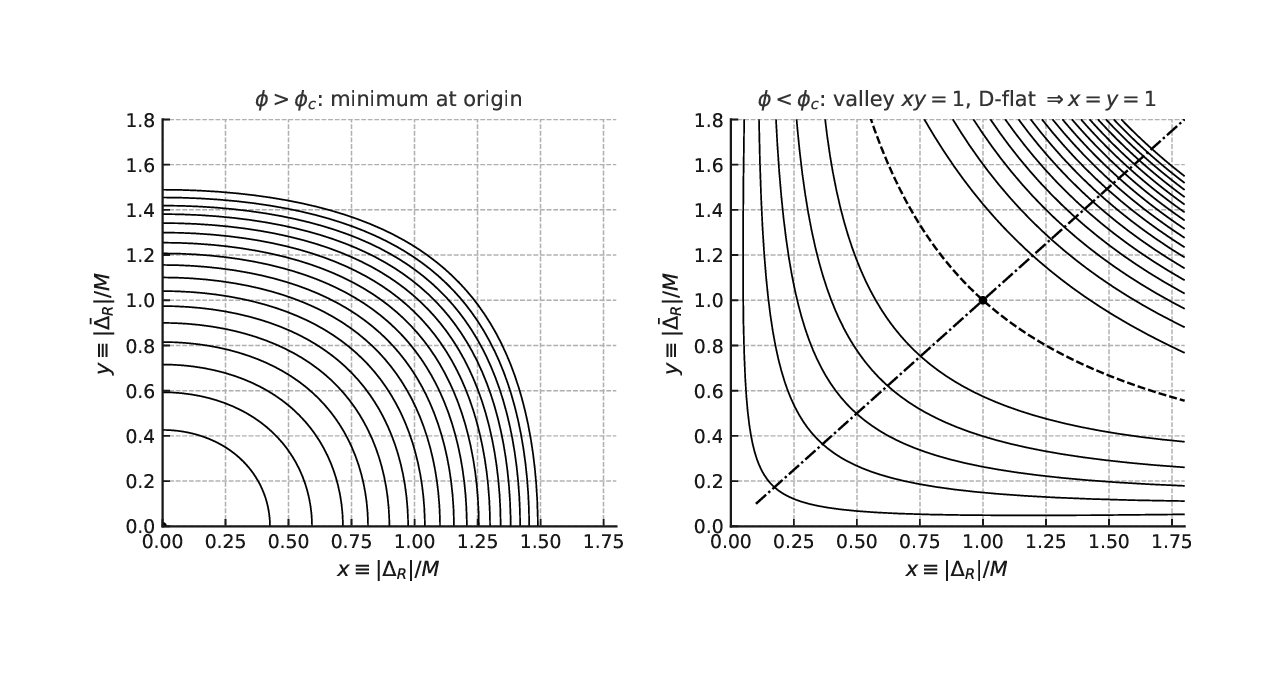}
 \caption{Schematic potential landscapes for the waterfall sector in terms of $x\equiv |\Delta_R|/M$ and $y\equiv |\overline{\Delta}_R|/M$. Left:for $\tilde{N}_1>\tilde{N}_{1c}$ the minimum is at $x=y=0$, so the inflationary valley lies at $\Delta_R=\overline{\Delta}_R=0$ with vacuum energy $V_0=\kappa^2M^4$. Right: for $\tilde{N}_1>\tilde{N}_{1c}$ a valley opens along the F-flat constraint $xy=1$ (dashed), and D-flatness $x=y$ (dash-dot) selects the true vacuum at $(x,y)=(1,1)$ where $V=0$.}
  \end{center}
  \label{fig:waterfall2}
\end{figure}
In this case, a condensation of the scalar field 
$\left<\tilde{N}_1 \right>$ causes the inflation 
and the successive reheating processes. 
Then the above potential drives the sneutrino 
$\tilde{N}_1$ (a hybrid inflation \cite{Linde, Dvali}) and it determines 
the inflaton (sneutrino) mass to be around 
$m_{\rm inf} \simeq \left(M_{\rm GUT} \, M_{R1} \right)^{1/2} 
\simeq 5.7 \times 10^{13}$ GeV. 
The hybrid inflation landscapes are given in Figs 1 and 2. 
The mass scale of the sneutrino, when identified as the inflaton, is appropriately set by the condition for coherent oscillation at the end of inflation, $H\approx \Gamma_{\tilde{N}_1}$ (where $H$ denotes the Hubble parameter). 
In our case, the potential comes from the superpotential
\be
W=\kappa S(\overline{\Delta}_R\Delta_R-M^2)+... \;,
\ee
where $S$ is some Pati-Salam single of ${\bf 210}$ representation (see Eq. \bref{W1}). Then the potential is
\be
V(\tilde{N}_1,\Delta_R,\overline{\Delta}_R)=\kappa^2\left(|\overline{\Delta}_R\Delta_R-M^2|^2+|\tilde{N}_1|^2(|\Delta_R|^2+|\overline{\Delta}_R|^2\right)+V_{loop}(\tilde{N}_1)\;.
\ee
During inflation:
\bea
\Delta_R&=&\overline{\Delta}_R=0;
V\approx V_o=\kappa^2M^4~\mbox{(a constant false -vacuum energy)} \;.
\eea
The slow-roll slope comes only from radiative corrections or SUGRA terms:
\be
V_{eff}(\tilde{N}_1)\approx V_0\left(1+\frac{\kappa^2}{8\pi^2}\ln\frac{|\tilde{N}_1|}{Q}\right) \;.
\ee
Hence
\be
A_s\approx \frac{V_0}{24\pi^2M_{Pl}^4\epsilon_*} \;, ~~\epsilon_*=\frac{M_{Pl}^2}{2}\left(\frac{V'}{V}\right)^2\approx \frac{M_{Pl}^2}{2}\left(\frac{\kappa^2}{8\pi^2\tilde{N}_1}\right)^2 \;,
\ee
where $A_s$ is the scalar curvature power spectrum  at the pivot scale $k_*$ (usually $0.05 \mbox{Mpc}^{-1}$)
The primordial density fluctuation is given by
\be
\frac{\delta T}{T}\approx \frac{1}{5}\sqrt{A_s}\approx 10^{-5} \;.
\ee
So the observed $10^{-5}$ directly reflect
\be
 A_s \approx 2\times 10^{-9} \;.
 \ee

The tree level sneutrino decay rate is given by 
\be
\Gamma_{\tilde{N}_1} 
\ \simeq\ \frac{1}{4 \pi} \left(Y_\nu Y_\nu^\dag \right)^{11} M_{R1}
\ \simeq\ 6.1 \times 10^{7} \; {\rm GeV} \;, 
\ee
where $Y_\nu$ is given in Eq. \bref{Ynu}, and 
we took the typical value of 
$\left(Y_\nu Y_\nu^\dag \right)^{11} \simeq 4.7 \times 10^{-3}$. 
Thus the reheating temperature in this model is given by \footnote{This value of the reheating temperature is a little bit high 
in considering the gravitino problem with the hadronic decay \cite{kohri}. 
To avoid such a difficulty, 
we just assume the heavy gravitino $m_{3/2} \gtrsim 100$ [TeV] motivated 
by the anomaly mediated SUSY breaking (AMSB) scenario \cite{anom1, anom2}.} 

\be
T_R \ =\ \left(\frac{45 \;M_{\rm P}^2 }{2 \pi^2 g_*} \right)^{1/4} 
\left(\Gamma_{\tilde{N}_1} \right)^{1/2}
\ \simeq\ 4.0 \times 10^{12} ~{\rm GeV} \;.
\ee
In general, $T_R\ll f_a=v_R$ is assumed to avoid thermal restoration of PQ symmetry.
Although the minimal SO(10) model is highly predictive, certain parameter ambiguities remain, allowing for the possibility of relatively low reheating temperatures, as will be discussed later.
After giving rise to the PQ symmetry breaking VEV of the Higgs,   
\be
\left<{\bf \bar{\Delta}}_R \right> \ \simeq\ 
8.3 \times 10^{12} \;{\rm GeV} 
\ee
from Eq.\bref{MR}, the argument of ${\bf \bar{\Delta}}_R$ can be regarded as 
the PQ field or an invisible axion, 
$a(x) \equiv f_a \times \left[\arg \left({\bf \bar{\Delta}}_R \right)
- \arg \left({\bf \Delta}_R \right) \right]$ 
with the decay constant 
$f_a = \left|\left<{\bf \bar{\Delta}}_R \right> 
\right| \ \simeq\ 8.3 \times 10^{12}$ GeV.

It should be noted that the gauged $B-L$ symmetry included 
in the SO(10) symmetry protects the sneutrinos from having large initial 
values along the existing $B-L$ flat direction. Therefore we can not 
incorporate the simple chaotic inflation scenario \cite{chaotic} 
into the SO(10) models, and we must use the hybrid inflation model. 
Recent WMAP data also supports the fact that multi-field hybrid 
inflation models are preferable to the single field chaotic inflation 
model \cite{WMAP1, WMAP2}. 
Even within the minimal SO(10) model, the numerical fit is not unique when the minimum of the $\chi^2$ analysis is slightly relaxed \cite{Fuku6}. The resulting values for four representative classes of parameters are summarized in Table 2. For the detailed structure of the Dirac neutrino Yukawa matrix $Y_\nu$, please refer to Ref.~\cite{Fuku6}.

\begin{table}[h]
\centering
\caption{Decay width $\Gamma_\phi$ and reheating temperature $T_{\rm RH}$ 
for the four benchmark cases of Fukuyama--Ichikawa--Mimura \cite{Fuku6}.
}
\label{tab:FIM_reheating}
\begin{tabular}{lcccc}
\hline
\textbf{Case (FIM 2016)} & 
$\boldsymbol{M_{R1}}$ [GeV] &
$\boldsymbol{(Y_\nu Y_\nu^\dagger)_{11}}$ &
$\boldsymbol{\Gamma_\phi}$ [GeV] &
$\boldsymbol{T_{\rm RH}}$ [GeV] \\
\hline
1.\ Type-II & $6.9\times10^{8}$  & $9.3\times10^{-5}$  & $5.1\times10^{3}$  & $5.1\times10^{10}$ \\
2.\ Type-II (best) & $2.4\times10^{10}$ & $2.2\times10^{-4}$ & $4.1\times10^{5}$  & $4.6\times10^{11}$ \\
3.\ Type-I (local) & $6.6\times10^{8}$  & $9.4\times10^{-5}$  & $4.9\times10^{3}$  & $5.1\times10^{10}$ \\
4.\ Type-I (global min.) & $6.8\times10^{9}$  & $1.3\times10^{-4}$  & $7.1\times10^{4}$  & $1.9\times10^{11}$ \\
\hline
\end{tabular}
\end{table}
Although the “best fit” in Ref.~\cite{Fuku6} corresponds to the global $\chi^2$ minimum in low-energy observables, a slightly sub-optimal solution can provide a more consistent picture once cosmological constraints—such as the reheating temperature, leptogenesis, and the PQ symmetry scale—are taken into account. Therefore, in the present framework, the physically favored solution does not necessarily coincide with the statistical $\chi^2$ minimum.


Finally we give some comments on the domain wall problem. From our scenario, 
$\Delta_R\neq 0$ appears after inflation, leaving the SM gauge group. Therefore, $Z_6$ symmetry remains for DFSZ model like ours.
Fortunately we can solve the domain wall problem via Lazarides-Shafi mechanism \cite{LS, Higaki} as follows.
The anomaly coefficient $A_{PQ}^{3}$ due to $U(1)_{PQ}-SU(3)_c-SU(3)_c$ vertex is
\be
A_{PQ}^3=\sum_{generation}Q_{16}(2T(3)+T(\overline{3})+T(\overline{3}))=3(2\times \frac{1}{2}+\frac{1}{2}+\frac{1}{2})Q_{16}=6
\ee
with $Q_{16}=1$ from Table 1. Here we have used quarks are composed of $q_L(3,2,1)+u^c(\overline{3},1,1)+d^c(\overline{3},1,1)$ in one generation with three generations.
Thus
\be
N_{DW}^{eff}=\frac{A_{PQ}^{(3)}}{p}=1 \;.
\ee
It has been shown that the PQ solution to the strong-CP problem—equivalently, the axion mechanism—can be embedded in the same framework as the right-handed neutrino sector. In particular, a complete correspondence between the axion and the Higgs field that generates the right-handed neutrino mass has been realized in the minimal SO(10) model. In this setting, the relative phase between $\bar{\Delta}_R$ and $\Delta_R$ can be identified with the axion. Consequently, the symmetry-breaking scales of PQ, $B\!-\!L$, and the right-handed neutrino mass are closely linked. 

Furthermore, taking the sneutrino to be the inflaton can be incorporated naturally in this framework. A sneutrino mass of order $10^{13}\,\mathrm{GeV}$ reproduces the observed amplitude of CMB density fluctuations. In this scenario the PQ and $B\!-\!L$ symmetries break after inflation ends, and the domain-wall problem is avoided via the Lazarides–Shafi mechanism.


\begin{thebibliography}{99}
\bibitem{PQ1}
R.D. Peccei, H.R. Quinn, 
Phys.\ Rev.\ Lett.\ {\bf 38}, 1440 (1977).
\bi{PQ2}
R.D. Peccei, H.R. Quinn, 
Phys.\ Rev.\ {\bf D 16}, 1791 (1977). 
\bibitem{WW1}
S. Weinberg, Phys.\ Rev.\ Lett.\ {\bf 40}, 223 (1978).
\bibitem{WW2} 
F. Wilczek, Phys.\ Rev.\ Lett.\ {\bf 40}, 279 (1978).

\bibitem{KSVZ1}
J. E. Kim, Phys.\ Rev.\ Lett.\ {\bf 43}, 103 (1979).
\bibitem{KSVZ2} 
M. Shifman, A. Vainstein and V. Zakharov, 
Nucl.\ Phys.\ {\bf B 166}, 493 (1980). 

\bibitem{ZDFS1}
A. Zhitnitsky, Sov.\ J.\ Nucl.\ Phys.\ {\bf 31}, 60 (1980).
\bibitem{ZDFS2}
M. Dine and W. Fishler, and M. Srednicki, 
Phys.\ Lett. B {\bf 104},199 (1981). 

\bi{Gellmann}
M. Gell-Mann, R.J. Oakes, and B. Renner, Phys. Rev. {\bf 175}, 2195 (1968).
\bi{Kim}
J. E. Kim and G. Carosi, Rev. Mod. Phys. {\bf 82}, 557 (2010).
\bi{Raffelt}
G. Raffelt and D. Seckel, Phys. Rev. Lett. {\bf 60}, 1793 (1988).
\bi{Janka}
H.T. Janka et al., Phys. Rept. {\bf 442}, 38 (2007).
\bibitem{Preskill}
J.~Preskill, M.~B.~Wise, and F.~Wilczek, Phys. Lett. B {\bf120}, 127 (1983).

\bibitem{Sikivie}
L.~F.~Abbott and P.~Sikivie, 
Phys. Lett. B \textbf{120}, 133 (1983).

\bibitem{Dine}
M.~Dine and W.~Fischler,  

Phys. Lett. B \textbf{120}, 137 (1983).

\bi{Svrcek}
P. Svrcek and E. Witten,  {\bf 06}, 051 (2006).
\bi{Visinelli}
L. Visinelli and S. Vagnozzi, Phys. Rev. {\bf D 99}, 063517 (2019).
\bibitem{Fuku7}
T. Fukuyama,  Int. J. Mod. Phys. {\bf A 27}, 1230015 (2012).
\bi{Lazarides}
G. Lazarides, Phys. Rev. {\bf D 25}, 2425 (1982).
\bi{Reiss}
D.B. Reiss, Phys. Lett. B {\bf 109}, 365 (1982).
\bi{Kikuchi}
T. Fukuyama and T. Kikuchi, JHEP {\bf 05}, 017 (2005).
\bi{Ernst}
A. Ernst, A. Ringwald, and C. Tamarit, JHEP {\bf 02}, 103 (2018).
\bi{Boucenna}
S. Boucenna, T. Ohlsson, and M. Pernow, Phys. Lett. B {\bf 792}, 251 (2019).
\bi{Hamada} 
Y. Hamada, M. Ibe, Y. Muramatsu, K. Oda, and N. Yokozaki, Eur. Phys. J. {\bf C80}, 482 (2020).
\bi{Babu2}
K.S. Babu, T. Fukuyama, S. Khan, S. Saad, JHEP {\bf 06}, 045 (2019). 
\bibitem{Fuku1}
K. Matsuda, Y. Koide and T. Fukuyama, 
Phys.\ Rev.\ {\bf D 64}, 053015 (2001).
\bibitem{Fuku11}
K. Matsuda, Y. Koide, T. Fukuyama and H. Nishiura, 
Phys.\ Rev.\ {\bf D 65}, 033008 (2002).
\bibitem{Babu}
K.S. Babu and R.N. Mohapatra, 
Phys.\ Rev.\ Lett.\ {\bf 70}, 2845 (1993). 
\bibitem{Fuku2}
T. Fukuyama and N. Okada, 
JHEP {\bf 0211}, 011 (2002).
\bi{Fuku3}
T.~Fukuyama, A.~Ilakovac, T.~Kikuchi, S.~Meljanac and N.~Okada,
J.\ Math.\ Phys.\  {\bf 46}, 033505 (2005).
\bibitem{Fuku4}
T.~Fukuyama, A.~Ilakovac, T.~Kikuchi, S.~Meljanac and N.~Okada, Eur. Phys. J. {\bf C42}, 191 (2005).
\bibitem{Fuku6}
T. Fukuyama, K. Ichikawa, and Y. Mimura, Phys. Rev. {\bf D 94}, 075018 (2016).
\bibitem{sneutrino}
H. Murayama, H. Suzuki, T. Yanagida, J. Yokoyama, 
Phys.\ Rev.\ Lett.\ {\bf 70}, 1912, (1993).
\bibitem{Ellis}
J.R. Ellis, M. Raidal, T. Yanagida, 
Phys.\ Lett.\ B{\bf 581} 9, (2004). 
\bibitem{Kallosh}
R. Kallosh, A. Linde, Q. Roest, and T. Wrase, JCAP, {\bf 11}, 046 (2016).
\bibitem{Antusch}
S. Antusch, AIP Conference Proceedings, {\bf 878}, 284 (2006).
\bibitem{Linde}
A. D. Linde, Phys. Rev. {\bf D 49}, 748 (1994).
\bibitem{Dvali}
G. R. Dvali, Q. Shafi, and R. Schaefer, Phys. Rev. Lett. {\bf 73}, 1886 (1994).
\bibitem{kohri}
M.~Kawasaki, K.~Kohri and T.~Moroi,
Phys. Rev. {\bf D 71}, 083502 (2005).
\bibitem{anom1}
L. Randall and R. Sundrum, 
Nucl.\ Phys.\ {\bf B 557}, 79 (1999).
\bibitem{anom2}
G.F. Giudice, M.A. Luty, H. Murayama and R. Rattazzi, 
JHEP {\bf 9812}, 027 (1998). 

\bibitem{chaotic}
A. Linde, Phys.\ Lett.\ B {\bf 129}, 177 (1983). 

\bibitem{WMAP1}
V. Barger, H.-S. Lee, D. Marfatia, 
Phys.\ Lett.\ B {\bf 565}, 33 (2003).
\bibitem{WMAP2}
W.H. Kinney, E.W. Kolb, A. Melchiorri, A. Riotto, 
Phys.\ Rev.\ {\bf D 69}, 103516 (2004). 


\bibitem{LS}
G. Lazarides and Q. Shafi, Phys. Lett. B {\bf 115}, 21 (1982).
\bibitem{Higaki}
C. Chattererjee, T. Higaki, and M. Niita, Phys. Rev. {\bf D 101}, 75026 (2020). 

\end{thebibliography}
\end{document}